\documentclass[aps,prl,reprint,superscriptaddress,nofootinbib,longbibliography,floatfix]{revtex4-2}
\usepackage{amsmath,amssymb,graphicx}
\usepackage{needspace}
\usepackage[patch=none]{microtype}
\usepackage[hidelinks]{hyperref}
\graphicspath{{figures/}}

\begin{document}

\title{Rapidly Spinning Primordial Black Holes from Confinement}
\author{Michael Zantedeschi}
\email{michael.zantedeschi@pi.infn.it}
\affiliation{INFN, Sezione di Pisa, Largo Bruno Pontecorvo 3, I-56127 Pisa, Italy}
\date{\today}

\begin{abstract}
Confinement can turn widely separated particles into rapidly spinning
primordial black holes. A flux tube pulls a heavy quark--antiquark pair
together, converting its stored energy into relativistic motion.
The same mechanism is possible in the magnetic dual, where a Higgs phase
confines magnetic flux into a string joining a monopole--antimonopole pair.
As the particles approach, cosmological tides and string bends can deflect
their motion and make them miss. These disturbances can originate during
inflation or be generated later. Black-hole formation then requires rare
close passages from a smooth, broad encounter
distribution. The resulting
natal spin density therefore rises linearly to extremality, with mean
dimensionless spin two thirds. Numerical string evolutions recover this
law. Pairs that miss can remain bound, radiate
and later collapse into lighter, near-extremal holes, sourcing a
stochastic gravitational-wave background. Although the abundance remains
model dependent, confinement leaves a unique spin signature that
gravitational-wave observations can test.
\end{abstract}

\maketitle
\setlength{\parskip}{3pt plus 1pt minus 0.5pt}

Confinement can bring widely separated particles together to form a
primordial black hole~\cite{Dvali:2021byy,Matsuda:2005ey}. Consider a heavy quark--antiquark pair separated by
inflation and connected by a tube of colour-electric flux. The tube stores
energy in proportion to the separation and releases it into the quarks'
motion as it contracts. A close encounter can therefore produce a black
hole far heavier than the pair's combined rest mass.
The same holds for monopoles, whose magnetic flux
can be confined by a Higgs transition into a vortex string joining a
monopole to an antimonopole. Depending on the tension and separation,
this mechanism can produce primordial black holes from ultralight to
solar masses and account for an order-one fraction of the dark matter~\cite{Dvali:2021byy}.

During contraction, the particles gain energy and can acquire orbital
angular momentum. Whether both enter the same compact region depends on
how the string bends and moves.
Cosmological tides give the particles transverse motion and bend the
string, while frozen bends inherited from inflation can later bring an
initially nonrotating pair to a spinning encounter. These sources can
provide a smooth encounter distribution, but they need not be the only
ones. When the encounter distribution is broad compared with the capture
region, most pairs miss.
Yet rare close encounters need not produce slowly rotating holes.
A pair can miss in two transverse directions, so larger offsets occupy
more area within the capture region. Since orbital angular momentum grows
with the offset at fixed energy, the leading natal law for the
dimensionless black-hole spin \(a_*\) is
\(p(a_*)=2a_*\), rising linearly to extremality, the maximum spin a black
hole can sustain, with mean dimensionless spin \(2/3\).

Numerical string evolution recovers this law for nearly complete capture
and for spinning compact cores formed from initially nonrotating strings,
with compensating angular momentum remaining in an outer string fold.
Bound pairs that initially miss can also lose energy and angular momentum
through gravitational radiation, forming lighter, near-extremal holes
and sourcing a stochastic gravitational-wave background.
A companion video shows these string evolutions~\cite{ConfinementVideo}.

For holes born from primordial
overdensities~\cite{Zeldovich:1967lct,Hawking:1971ei,Carr:1974nx,Carr:1975qj},
however, the spin depends on the cosmological
background~\cite{Escriva:2022duf,Green:2020jor,Carr:2026hot}.
Radiation-era collapse generally produces slowly rotating
holes~\cite{Harada:2020pzb,Banerjee:2024spin}. Matter domination can allow
more rapidly rotating holes~\cite{Harada:2017MDspin,Saito:2024MDspin},
although this expectation has recently been challenged by calculations
predicting small spins for weak density fluctuations~\cite{Ye:2025MDspin}.
With confinement, tension drives the
contraction, so the linear spin law applies during both radiation and
matter domination. At stellar masses, gravitational-wave
measurements of spin and its mass dependence can therefore test this
prediction and distinguish formation histories that masses alone leave
unresolved~\cite{Fishbach:2017dwv,Gerosa:2017kvu,DeLuca:2020spinevolution,DeLuca:2020LVdata,Berti:2026massspin}.

\begin{figure*}[t]
 \centering
 \includegraphics[width=\textwidth]{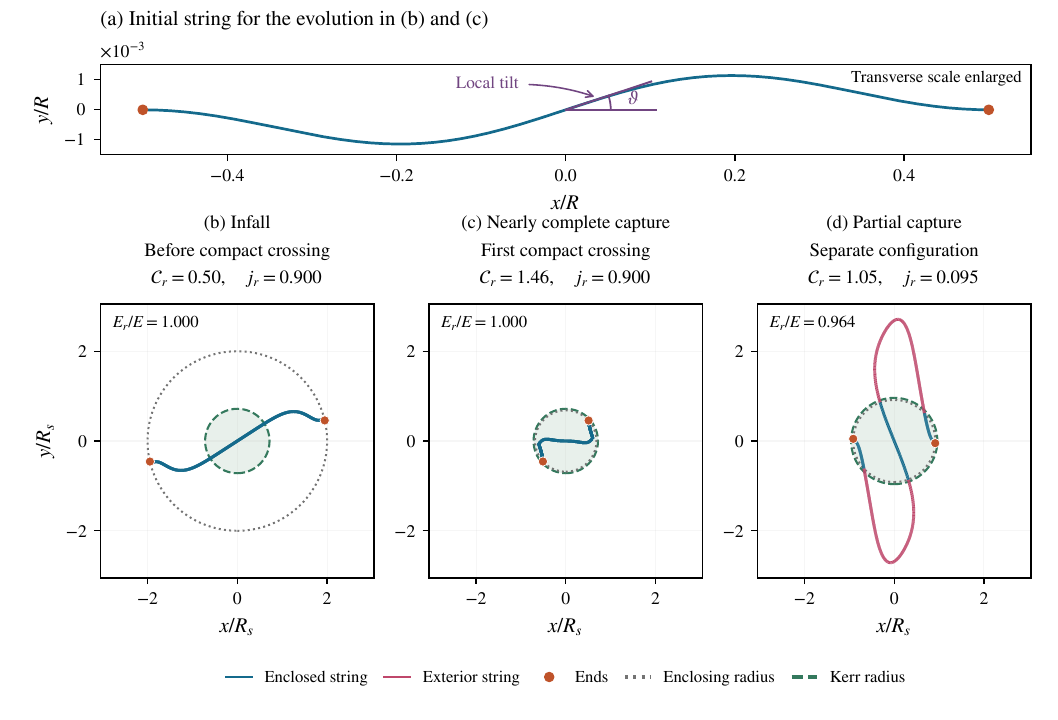}
 \caption{String contraction and capture.
 (a) Initial balanced-S shape, with enlarged transverse scale and maximum
 tilt \(\vartheta\) marked in violet. (b,c) Nearly complete capture.
 (d) Partial capture from an initially nonrotating string.
 Orange marks particles, blue enclosed string, pink exterior string.
 The dotted circle encloses energy \(E_r\), with compactness
 \(\mathcal C_r\) and spin \(j_r\) defined in Eq.~\eqref{eq:enclosed}.
 The dashed circle marks the Kerr radius of Eq.~\eqref{eq:criterion}.}
 \label{fig:dynamics}
\end{figure*}

\medskip
\textbf{Contraction.}
We consider two particles of mass \(m\),
joined by an isolated thin flux tube of tension \(\mu\) and initial length
scale \(R\). Inflation can separate the pair
before or after the tube forms~\cite{Dvali:2021byy}.
At horizon re-entry, the separation is of order the horizon size and
tension pulls the particles together. The tube must survive infall:
\(\mu\lesssim m^2\) suppresses pair-production tunneling~\cite{Dvali:2021byy,PreskillVilenkin1993},
while radiative and plasma losses must remain small.

In natural units, the gravitational coupling, incoming energy, and
gravitational radius are
\begin{equation}
 \alpha_{\rm gr}\equiv G\mu,\qquad E\simeq\mu R,\qquad R_s\equiv2GE.
 \label{eq:scales}
\end{equation}
Small coupling keeps gravity weak during the contraction, while
stored energy much larger than the pair's rest energy makes the particles
relativistic. The gravitational radius must exceed the microscopic string
and particle widths, so that their finite size does not prevent the
encounter energy from fitting inside their gravitational horizon.

\medskip
\needspace{3\baselineskip}
\textbf{Spin from close encounters.}
By closest approach, the particles carry most of the energy in
relativistic motion. We can therefore relate the captured spin to their
incoming trajectories. In their center-of-momentum frame, let
\(E_{\rm pair}\) be their combined relativistic energy immediately before
capture, and \(b=|\mathbf b|\) the transverse
separation of their incoming trajectories. During a compact encounter
lasting \(\sim GE_{\rm pair}\), tension changes the particle momenta by only a
fraction \(\sim\alpha_{\rm gr}\). The trajectories are therefore nearly
straight locally, even when the connecting string is bent. To leading
order, capture inherits their energy and orbital angular momentum:
\begin{equation}
 J_{\rm pair,in}\simeq\frac{E_{\rm pair}b}{2},\qquad
 a_*\simeq\frac{J_{\rm pair,in}}{GE_{\rm pair}^2}
      \simeq\frac{b}{2GE_{\rm pair}}.
 \label{eq:encounter}
\end{equation}
Here \(a_*\) is the dimensionless black-hole spin and
\(M_{\rm BH}\simeq E_{\rm pair}\). Taking the extremality limit as the capture
boundary selects \(b\lesssim2GE_{\rm pair}\).

At fixed encounter energy and direction, \(\mathbf b\) has two
components. A smooth, broad distribution is nearly uniform across the
capture region, so the probability in a ring is proportional to its area,
\(2\pi b\,db\).
Through Eq.~\eqref{eq:encounter}, this gives the leading natal spin law
\begin{equation}
 p(a_*)=2a_*,\qquad 0\leq a_*\leq1.
 \label{eq:spinlaw}
\end{equation}
Its mean and standard deviation are
\begin{equation}
 \langle a_*\rangle=\frac{2}{3},\qquad
 \sigma_{a_*}=\frac{1}{3\sqrt{2}}\simeq0.24.
 \label{eq:spinmoments}
\end{equation}
The density gradient averages away around the ring, leaving corrections
quadratic in the ratio of capture radius to distribution width.
Tides and bending determine how often capture occurs, while the local
geometry of successful encounters fixes the leading spin distribution.

\medskip
\textbf{Angular momentum.}
We begin with nearly complete capture, where the hole inherits the
incoming energy and angular momentum before appreciable radiation escapes:
\begin{equation}
  J_K\equiv GE^2,\qquad j\equiv\frac{|\mathbf J|}{J_K},\qquad
  a_*\simeq j.
 \label{eq:inheritance}
\end{equation}
Here \(\mathbf J\) is the total angular momentum of the string and both
endpoints in their center-of-momentum frame, \(J_K\) the maximum Kerr value
for the incoming energy.

For a sphere of radius \(r\) enclosing the whole source, \(J\leq Er\)
bounds its compactness by \(2GE/r\leq2/j\). Thus \(j\gg1\) prevents
nearly complete collapse unless angular momentum escapes. Partial capture
can still occur if exterior string carries the excess angular momentum.

One route to rare encounters is a broad distribution of the initial
angular momentum,
\begin{equation}
 J_K\ll\sigma_J,
 \label{eq:rare}
\end{equation}
where \(\sigma_J\) is the characteristic width of each component of the
total transverse angular momentum. For nearly straight pairs these lie
perpendicular to the pair axis.

\textit{Tides.} Cosmological tides provide one
contribution~\cite{CatelanTheuns1996}. The tide can originate from
inflationary scalar perturbations acting gravitationally on the pair
during contraction. Around horizon entry, a Newtonian potential perturbation of
amplitude \(\Phi\) across the pair produces relative transverse speed of
the same order. A uniform field cannot produce this relative motion.
The long lever arm then gives
\begin{equation}
 \sigma_J^{\rm tide}\sim\Phi ER\sim\Phi\mu R^2.
 \label{eq:tidalwidth}
\end{equation}
The Kerr scale instead involves the much smaller gravitational radius.
With tides alone, Eqs.~\eqref{eq:inheritance} and \eqref{eq:tidalwidth} give
\(J_K/\sigma_J\sim\alpha_{\rm gr}/\Phi\). Corrections to
Eq.~\eqref{eq:spinlaw} from variation of the parent density are then of
order \((J_K/\sigma_J)^2\).
Tides act on the particles and do not require the string to have
existed during inflation. The differential force also launches transverse
disturbances along the string.

\medskip
\textbf{Does shape matter?}
Since tides also bend the string, we must ask whether the energy can still
reach a compact region. We describe the initial deformation by
\begin{equation}
 \mathbf X_\perp(\tau=0,\sigma)=\vartheta R\,\mathbf f(\sigma/R),
 \qquad \max|\mathbf f'|=1,
 \label{eq:shape}
\end{equation}
Here \(\mathbf X_\perp\) is the transverse displacement at time \(\tau\),
while \(\sigma\) approximates longitudinal distance for small bends and
speeds. The dimensionless profile \(\mathbf f\) is normalized so that
\(\vartheta\) is the maximum initial tilt, shown in
Fig.~\ref{fig:dynamics}(a). Its square estimates the mean-square tilt up
to a profile-dependent factor.

Bending gives the string an excess length of order \(\vartheta^2R\)
relative to its longitudinal separation. After the leading transverse bend
refocuses, this sets the residual extent
\(r_{\rm bend}\sim\vartheta^2R\) at closest approach, as the numerical
evolutions confirm. Comparing it
with the gravitational radius in Eq.~\eqref{eq:scales} gives
{
\begin{equation}
 \frac{r_{\rm bend}}{R_s}
 \sim\frac{\left\langle|\partial_\sigma\mathbf X_\perp|^2\right\rangle}
           {\alpha_{\rm gr}}
 \sim\frac{\vartheta^2}{\alpha_{\rm gr}}\ll1.
\label{eq:window}
\end{equation}
}
The brackets average the squared tilt along the initial string, including
shorter-wavelength bends. Each bend is weighted by the fraction of string
it occupies. For sub-Kerr encounters this
allows nearly complete enclosure without perfect straightening.

When \(\vartheta^2\) becomes comparable to \(\alpha_{\rm gr}\), enclosure
becomes sensitive to the profile. Larger bends can leave an outer string
fold during capture, as in Fig.~\ref{fig:dynamics}(d).

For a coherent tide with \(\vartheta\sim\Phi\), the window
\(\Phi^2\ll\alpha_{\rm gr}\ll\Phi\) makes close encounters rare while
keeping the residual bend small enough for nearly complete capture.

A string already present during inflation can also acquire transverse fluctuations
with pair-scale tilt \(\vartheta\sim H_{\rm inf}/\sqrt{\mu}\)~\cite{GarrigaVilenkin1993,StephanOtto2004},
where \(H_{\rm inf}\) is the inflationary Hubble rate.
Avoiding strong exponential suppression of string nucleation during
inflation requires \(\mu\) not much larger than
\(H_{\rm inf}^2\)~\cite{Basu:1991ig,GarrigaVilenkin1993,Dvali:2021byy}, giving
\(\vartheta\sim \mathcal{O}(1)\), outside the focusing condition of
Eq.~\eqref{eq:window}. Such order-unity bends make close encounters rare
without changing the geometrical argument for the natal spin law. They
instead lead to a different capture probability, discussed below.
Appendix~\ref{sup:selection} gives the slow-roll comparison and
shorter-wavelength estimates.

After inflation, thermal gauge
fields~\cite{HindmarshRajantie2000} and bubble
collisions~\cite{KibbleVilenkin1995} can set the initial configuration of
Higgs strings. Plasma flows from a first-order
transition~\cite{HindmarshHijazi2019} can bend them further when the string
or its endpoints couple to the bath.
Tides remain operative in either case,
while shape alone can control encounters when tidal torque is negligible.

These sources can be compared through the transverse encounter-offset
width \(\sigma_b\). For nearly complete capture,
Eq.~\eqref{eq:encounter} gives \(\sigma_b\simeq2\sigma_J/E\), hence
\(\sigma_b\sim\Phi R\) for tidal motion. For initially stationary bent
strings, the numerical ensembles instead give
\(\sigma_b\sim\vartheta^2R\).

\medskip
\textbf{Capture probability.}
The same encounter geometry gives a capture probability quadratic in the
ratio of capture radius to typical impact parameter.
Writing this first-encounter probability as \(P_{\rm prompt}\), nearly
complete capture gives
\begin{equation}
 P_{\rm prompt}\sim\left(\frac{J_K}{\sigma_J}\right)^2.
 \label{eq:probability}
\end{equation}
The abundance also depends on pair density and separations.

For the strongly bent inflationary example with \(\vartheta\sim \mathcal{O}(1)\),
the encounter offsets instead spread over the pair separation \(R\).
The capture area relative to this spread gives the parametric estimate
\((GM_{\rm BH}/R)^2\sim\alpha_{\rm gr}^{\,2}\) for the collapse probability.

\medskip
\textbf{String dynamics.}
To check whether contraction brings energy and angular momentum into
the same compact region, we evolve subhorizon Nambu--Goto strings with
massive endpoints in flat spacetime~\cite{Martin:1996cp,Siemens2001}.
The companion video illustrates prompt capture and repeated encounters~\cite{ConfinementVideo}.

For a sphere of radius \(r\) containing both endpoints, with energy
\(E_r\) and angular momentum \(\mathbf J_r\) about its center, the
compactness and local Kerr ratio corresponding to
Eqs.~\eqref{eq:scales} and \eqref{eq:inheritance} are
\begin{equation}
 \mathcal C_r\equiv\frac{2GE_r}{r},\qquad
 j_r\equiv\frac{|\mathbf J_r|}{GE_r^2}.
 \label{eq:enclosed}
\end{equation}
We combine them into the condition
\begin{equation}
 \mathcal C_K(r)\equiv\frac{\mathcal C_r}{2}
       \left(1+\sqrt{1-j_r^2}\right)\geq1,
 \qquad j_r\leq1.
 \label{eq:criterion}
\end{equation}
We adopt Eq.~\eqref{eq:criterion} as a Kerr-motivated collapse criterion
during inward motion: the enclosed energy and angular momentum fit within
their Kerr-radius scale. Any exterior string is excluded from both
quantities. An encounter region slightly above its own Kerr bound
could collapse by radiating angular momentum. This uncounted route
enhances the high-spin tail.

To separate bending from initial motion, we evolve bent strings both
with smooth initial transverse velocities and from rest
(Appendix~\ref{sup:dynamics}). For small planar bends satisfying
Eq.~\eqref{eq:window}, the first sub-Kerr encounter captures nearly all
the energy and angular momentum. Bending out of the plane mainly changes
the distribution near maximal spin. For strings starting at rest, the
straight-string limit instead gives nearly complete capture with vanishing spin.

Stronger bending can still allow capture because the endpoints carry
most of the energy in relativistic motion. They can therefore form a
compact core while an exterior string fold remains outside, as in
Fig.~\ref{fig:dynamics}(d). The core inherits the local encounter spin
of Eq.~\eqref{eq:encounter}. For an initially stationary string, the core's
angular momentum is balanced by that of the exterior string fold. The later spin
then depends on what happens to this string: complete absorption
without angular momentum escaping would leave a nonrotating hole.
Black-hole--string dynamics allows angular-momentum exchange and
reconnection-driven ejection~\cite{Xing2021,Deng2023}, while vortex
flux trapping has also been studied in the massive-photon
setting~\cite{DvaliOsmanovZantedeschi2025}.

Once the typical spread of encounter offsets exceeds the capture scale,
both the velocity and stationary ensembles recover the linear natal
spin law. Figure~\ref{sup:fig:inheritance}
shows the velocity ensembles, while Appendix~\ref{sup:focusing}
compares the stationary-string distributions with the area law.
Figure~\ref{sup:fig:shapeonly} illustrates one such encounter.
Detailed numerical dynamics will appear in a longer paper.

\medskip
\textbf{Delayed collapse.}
If a pair misses its first encounter and the string survives, tension
draws it together again.

Initially super-Kerr pairs that radiate only gravitationally can collapse
on a later encounter if their angular momentum falls fast enough to cross
the extremality bound, which itself decreases as energy is lost. The
relative losses depend on the string
shape and motion~\cite{Martin:1996cp,Babichev2005}. Since radiation changes
the pair only slightly per oscillation, it approaches the capture threshold
gradually and first reaches it close to extremality, after some energy
has already escaped. Numerical evolutions confirm this route to lighter,
near-extremal captured cores (Appendix~\ref{sup:radiation}).

Shape evolution provides another route to later capture, even without
radiation. An initially nonrotating string can redistribute angular
momentum between its parts while keeping the total zero.
For smooth encounter distributions, successive captures retain
the spin law~\eqref{eq:spinlaw}. They need not accumulate near extremality.

Over many encounters, however, even interactions negligible during one
infall can outpace gravitational losses~\cite{Chitose2025}. Plasma drag, torques and nongravitational
radiation can therefore change both the delayed population and its
gravitational-wave emission.

\medskip
\textbf{Possible realizations.}
Brane inflation provides a natural setting for particle--string
configurations~\cite{DvaliTye1999,DvaliThermalBrane1999,DvaliShafiSolganik2001}.
Wrapped branes can appear as massive particles in four dimensions, with
connecting strings produced when brane annihilation ends
inflation~\cite{SarangiTye2002,DvaliVilenkin2004,CopelandMyersPolchinski2004,Dvali:2021byy}.
Related configurations and their collapse also arise through the
cosmological slingshot effect~\cite{DvaliVilenkin2003,BachmaierSlingshot2023,BachmaierSlingshot2026}.

The mechanism can also operate in QCD if its early-Universe parameters
differ sufficiently from their present values~\cite{Dvali:2021byy,Dvali:1995ce}.
All string-breaking quarks must remain heavy relative to confinement
during contraction, while the
endpoints remain light compared with the stored energy and the bath allows
confinement.
Plasma and radiative losses
must leave the energy and selected angular momentum nearly unchanged
during the first infall~\cite{BraatenThoma1991,PeigneSmilga2009,BerezinskyMartinVilenkin1997}.
Appendix~\ref{sup:plasma} gives these loss conditions.

\textit{Abundance.} For nearly complete capture of pairs entering the horizon at cosmological time \(t\),
\(M_{\rm BH}\sim\mu t\). Following Ref.~\cite{Dvali:2021byy}, radiation-era
formation with order-one suitable pairs per horizon gives a present prompt dark-matter fraction
\begin{equation}
 \begin{aligned}
 f_{\rm PBH}^{\rm prompt}\simeq10^{-3}
 \left(\frac{P_{\rm prompt}}{10^{-3}}\right)
 \left(\frac{\alpha_{\rm gr}}{10^{-6}}\right)^{\frac{3}{2}}
 \left(\frac{M_\odot}{M_{\rm BH}}\right)^{\frac{1}{2}}.
 \end{aligned}
 \label{eq:abundance}
\end{equation}
The estimate assumes a standard thermal history, conserved entropy and
negligible later mass loss, suppressing order-one
geometric and relativistic degrees-of-freedom factors.
For shape-dominated encounters, successive
passages that explore fresh offsets can enhance the capture
probability by a factor parametrically as large as
\(1/\alpha_{\rm gr}\) before gravitational radiation removes a sizeable
fraction of the pair energy~\cite{Martin:1996cp}. For the inflationary example
with \(\vartheta\sim\mathcal O(1)\), repeated encounters can enhance the
abundance by weakening the capture suppression from
\(\alpha_{\rm gr}^2\) to \(\alpha_{\rm gr}\).
The gravitational-wave signal from these shape-driven encounters
is left for future investigation.

For illustration, consider a tide-dominated confining sector.
The CMB normalization motivates \(\Phi\sim10^{-5}\) as a reference
amplitude~\cite{Planck:2018vyg}, although its value on the smaller
scales probed here can differ. With
\(\alpha_{\rm gr}\sim2.4\times10^{-8}\) and
\(P_{\rm prompt}\sim6\times10^{-6}\),
Eq.~\eqref{eq:abundance} allows holes near
\(10^{18}\,\mathrm g\) to supply order-one dark matter.
Alternatively, \(\alpha_{\rm gr}\sim6\times10^{-7}\) and
\(P_{\rm prompt}\sim3.6\times10^{-3}\) give a fraction of order
\(10^{-3}\) in holes near \(3M_\odot\), relevant for
gravitational-wave spin measurements.

\begin{figure}[!t]
 \centering
 \includegraphics[width=\columnwidth]{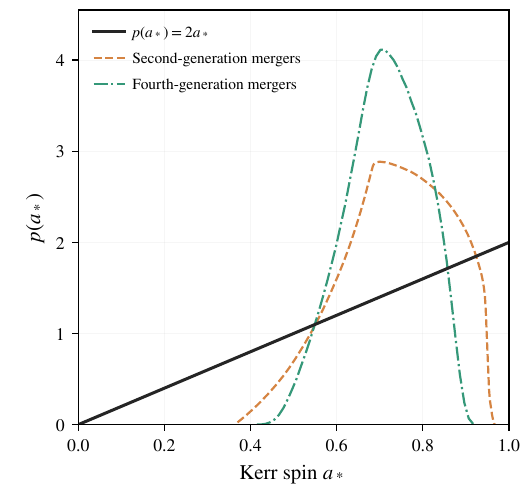}
 \caption{Confinement spin law \(p(a_*)=2a_*\) and equal-mass hierarchical
 mergers with isotropic spin directions, redrawn from
 Ref.~\cite{Fishbach:2017dwv}. Orange: second generation from initial
 spins 0.99. Green: fourth-generation attractor.}
 \label{fig:spin}
\end{figure}

\medskip
\textbf{Signatures.}
The linear natal spin law differs from the peaked distributions of
representative repeated-merger populations~\cite{Fishbach:2017dwv,Gerosa:2017kvu}, even when their means are similar,
as Fig.~\ref{fig:spin} shows. Radiatively delayed holes can add a concentration near
maximal spin and are lighter than their seeds, although a broad mass
spectrum allows both populations to overlap at the same observed mass.
Testing this prediction requires following the natal population into
binaries and accounting for subsequent accretion, spin orientations and
detector selection~\cite{DeLuca:2020spinevolution,LVK:2026SSM}.

PBHs formed through confinement are also expected to be strongly affected
by memory
burden~\cite{DvaliMemory2018,DvaliEisemannMichelZell2019,DvaliEisemannMichelZell2020,DvaliValbuenaZantedeschi2024,DvaliZantedeschiZellTransition2025,DondariniEtAl2025}.
Such confinement-driven formation is expected to produce a maximal memory
load~\cite{DvaliZantedeschiZell2026}.
The swift realization of this effect can then leave sizeable imprints on
merger dynamics and ringdown
frequencies~\cite{DvaliSwiftMemory2025,DvaliZantedeschiZell2026}.

\needspace{4\baselineskip}
Another proposal that this high-spin population could test is that rapidly
rotating black holes support
vorticity~\cite{DvaliKuhnelZantedeschi2022}, with consequences for the
radiation emitted in their mergers~\cite{DvaliVortexMergers2024}.

The energy radiated before delayed capture also supplies a stochastic
gravitational-wave background~\cite{Phinney:2001di}. For illustration,
we assume capture of pairs initially with \(1<j<1.5\) and a radiated energy
fraction of order ten percent. The area law then gives comparable prompt
and delayed abundances (Appendix~\ref{sup:background}).

For steady power \(50G\mu^2\)~\cite{Martin:1996cp,BuchmullerDomckeSchmitz2021} and emission ending
well after horizon entry during radiation domination,
Eq.~\eqref{eq:abundance} gives the frequency-integrated background
\begin{equation}
 \Omega_{\rm GW}\simeq10^{-13}
 \left(\frac{P_{\rm prompt}}{10^{-3}}\right)
 \left(\frac{\alpha_{\rm gr}}{10^{-7}}\right)^{1/2}.
\label{eq:gw-background}
\end{equation}
The leading mass dependence cancels between abundance and redshift.

The corresponding frequency observed today is
\(f\sim0.15\,\mathrm{Hz}\,
\left({M_{\rm BH}}/{10^{18}\,\mathrm{g}}\right)^{-1/2}\),
up to thermal factors.
At fixed mass, weaker tension gives a longer string and slower
oscillations, but also later emission and less redshift. In the radiation
era, these effects cancel at leading order.
The asteroid and stellar examples lie, respectively, in the DECIGO/BBO
target band~\cite{Kawamura:2021DECIGO,Corbin:2006BBO} and at a few nanohertz
in the pulsar-timing band~\cite{Agazie:2023NANOGrav}.
Both emit higher frequency harmonics. A complete characterization of the spectrum
and gravitational-wave signal is left for future work.

Confinement leaves its imprint in both black-hole spins and gravitational
waves. Rare close encounters give the general linear natal spin law,
while radiation from pairs that initially miss produces a stochastic
background. Their later collapse can add lighter, near-extremal holes,
linking the high-spin excess to the gravitational-wave signal.

\begin{acknowledgments}
I thank L.~Visinelli and J.~S.~V.~Bermudez for reading an early version
of the manuscript, G.~Dvali for discussions on this topic over the years,
and A.~Iovino for suggesting useful references and reading the manuscript.
\end{acknowledgments}

\setlength{\parskip}{0pt}
\bibliography{references}

\clearpage
\onecolumngrid
\begin{center}
 {\large\bfseries Supplemental material}\\[4pt]
 Rapidly Spinning Primordial Black Holes from Confinement
\end{center}
\twocolumngrid
\setlength{\parskip}{3pt plus 1pt minus 0.5pt}
\appendix
\setcounter{secnumdepth}{2}
\setcounter{figure}{0}
\renewcommand{\thefigure}{S\arabic{figure}}
\renewcommand{\theHfigure}{supp.\arabic{figure}}

\section{String evolution and capture}
\label{sup:dynamics}

We use a Nambu--Goto string with two particles of mass \(m\) at its
ends~\cite{Martin:1996cp,Siemens2001},
\begin{equation}
 S=-\mu\int d^2\zeta\sqrt{-\gamma}-m\sum_{a=L,R}\int ds_a.
 \label{sup:action}
\end{equation}
Here \(\zeta\) are world-sheet coordinates, \(\gamma\) is the determinant
of the induced metric, and \(ds_a\) is particle proper time. In temporal
conformal gauge the position \(\mathbf X(\tau,\sigma)\) obeys
\begin{equation}
 \ddot{\mathbf X}=\mathbf X'',\qquad
 \dot{\mathbf X}\cdot\mathbf X'=0,\qquad
 \dot{\mathbf X}^{\,2}+\mathbf X'^{\,2}=1.
 \label{sup:bulk}
\end{equation}
Dots and primes denote derivatives with respect to source-frame time
\(\tau\) and string coordinate \(\sigma\). At each end, tension gives
\begin{equation}
 \ddot{\mathbf x}_{L,R}
 =\pm\frac{\mu}{m}\gamma_{L,R}^{-3}
 \left.\frac{\mathbf X'}{|\mathbf X'|^2}\right|_{L,R},\qquad
 \mathbf v_a=\dot{\mathbf X}+\dot\sigma_a\mathbf X',
 \label{sup:boundary}
\end{equation}
where \(\mathbf x_a=\mathbf X(\tau,\sigma_a(\tau))\),
\(\gamma_a=(1-|\mathbf v_a|^2)^{-1/2}\), and the upper sign is for
the left endpoint~\cite{Siemens2001}. We propagate waves along the
string and integrate these endpoint equations. As the particles gain
energy, the coordinate interval occupied by the string shrinks.

The charges include both the string and its endpoints:
\begin{align}
 E&=\mu\int d\sigma+\sum_a m\gamma_a,\nonumber\\
 \mathbf P&=\mu\int d\sigma\,\dot{\mathbf X}
                  +\sum_a m\gamma_a\mathbf v_a,\nonumber\\
 \mathbf J&=\mu\int d\sigma\,\mathbf X\times\dot{\mathbf X}
                  +\sum_a m\gamma_a\mathbf x_a\times\mathbf v_a.
 \label{sup:charges}
\end{align}
All integrals run between \(\sigma_L\) and \(\sigma_R\). We work in the
total center-of-momentum frame. Restricting the energy and angular
momentum to a sphere containing both particles gives
Eq.~\eqref{eq:enclosed}, with angular momentum measured about the sphere's
center. Capture is recorded at the first inward crossing of
Eq.~\eqref{eq:criterion}. Concentration
of relativistic energy is also the basis of trapped-surface collision
calculations and string-loop collapse
studies~\cite{EardleyGiddings2002,Yoshino:2005hi,Helfer:2018qgv}.

\textit{Shape and motion.}
We denote the component of \(\mathbf X\) along the initial pair axis
by \(X_\parallel\). For a string initially at rest, its transverse shape
in Eq.~\eqref{eq:shape} and the unit-tangent constraint in
Eq.~\eqref{sup:bulk} fix
\(\partial_\sigma X_\parallel=\sqrt{1-\vartheta^2|\mathbf f'|^2}\).
To vary motion independently, we use
\begin{equation}
 \mathbf X'=\cos q\,\widehat{\mathbf t},\qquad
 \dot{\mathbf X}=\sin q\,\widehat{\mathbf v},\qquad
 \widehat{\mathbf t}\cdot\widehat{\mathbf v}=0,
 \label{sup:velocity}
\end{equation}
where the hats denote unit vectors and \(q(\sigma)\) sets the speed.
The rotating planar families use \(q=\arcsin(v_{\rm end})\sigma/\sigma_h\),
with initial endpoint speed \(v_{\rm end}\) and half-interval \(\sigma_h\).
Planar bends include the straight, balanced-S, bow and double-bow
profiles. To make nonplanar bends, we add four or eight harmonics to the
transverse tangent. Their coefficients are drawn at random from
zero-mean Gaussian distributions, with the same standard deviation for
all harmonics or one decreasing as \(1/n\), where \(n\) is the harmonic
number. We then align the initial endpoints and rescale each shape so
that its maximum transverse tangent is \(\vartheta\).
The mean-square tilt is calculated from that shape.

The main ensembles use \(\mu R/m=18\). Tidal checks also span
\(6,60,180\). Stationary shape-only ensembles set every velocity to zero
and hence have exactly
\(\mathbf J=0\). Any spin acquired by a captured core is balanced by
the exterior string.

\section{What the capture tests establish}
\label{sup:focusing}

\textit{Nearly complete capture.}
As discussed around Eq.~\eqref{eq:window}, a small bend adds a length
of order \(\vartheta^2R\). The simulations find the same quadratic
scaling for the size of the string at closest approach. This refers to
the space occupied by the whole string, including any remaining fold.
The particles themselves can pass much closer together.

At \(\alpha_{\rm gr}=10^{-4},3\times10^{-4},9\times10^{-4}\), with
\(\vartheta=\alpha_{\rm gr}^{2/3}\), all 72 planar sub-Kerr runs capture.
They enclose more than 99 percent of the energy and inherit the incoming
spin to within one percent. In 2016 nonplanar runs, bending mainly
affects encounters closest to extremality, where some shapes no longer
capture. Changing the coupling moves this upper edge slightly but leaves
the broad rising distribution intact: its mean spin remains
0.662--0.665, close to \(2/3\). Shorter-wavelength bends have a similarly
small effect. When the total mean-square tilt is at most
\(0.3\alpha_{\rm gr}\), they remove at most four percent of the
population that would otherwise capture.

\textit{Partial capture.}
To test the local encounter relation in Eq.~\eqref{eq:encounter}, we
evaluate the pair energy and orbital spin before capture:
\begin{equation}
 E_{\rm pair}^2=(E_L+E_R)^2-|\mathbf p_L+\mathbf p_R|^2,\qquad
 j_{\rm pair,in}=\frac{J_{\rm pair,in}}{GE_{\rm pair}^2},
 \label{sup:pairspin}
\end{equation}
with orbital angular momentum evaluated in the pair's center-of-momentum
frame before capture, at separation about \(8R_s\).

\begin{figure*}[t]
 \centering
 \includegraphics[width=\textwidth]{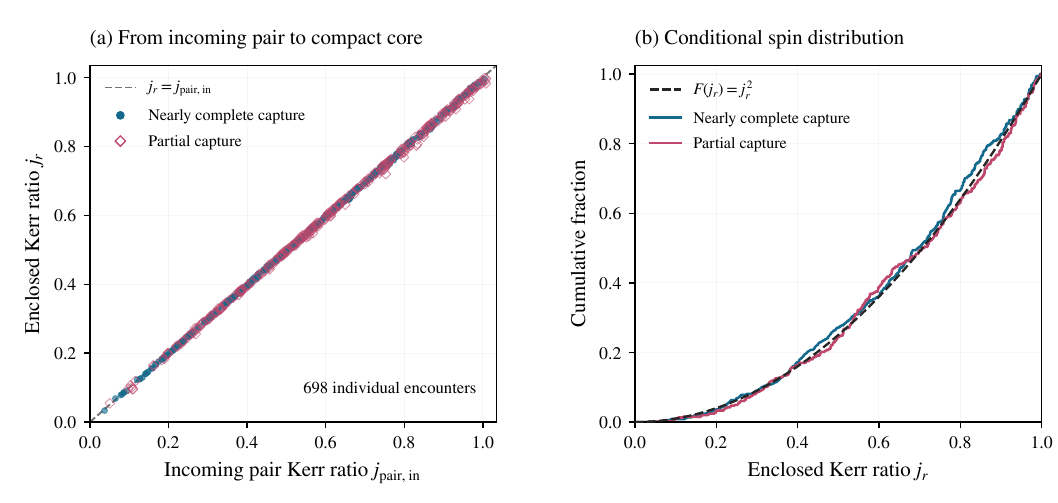}
 \caption{Nearly complete capture (blue, \(\vartheta=0.01\)) and
 partial capture (pink, \(\vartheta=0.12\)).
 (a) Enclosed versus incoming pair spin, with equality dashed.
 (b) Weighted cumulative distributions among captured pairs, compared
 with \(F(j_r)=j_r^2\).}
 \label{sup:fig:inheritance}
\end{figure*}

Figure~\ref{sup:fig:inheritance} compares nearly complete and partial
capture from smooth initial transverse velocities. Twelve nonplanar
shapes, three bend amplitudes and 48 velocity draws give 1728
trajectories. The target velocity distribution is Gaussian, centered at
zero, with standard deviation \(\vartheta\) in each transverse component
and no preferred transverse direction. The two samples shown contain
400 and 298 captures, with typical enclosed energy fractions of about
99 and 82 percent. In both, the core spin closely follows the incoming
pair spin. With weights corresponding to this Gaussian distribution,
the mean spins are 0.656 and 0.669, in agreement with
Eq.~\eqref{eq:spinmoments}. Bending shifts the motions that lead to
capture. When the shift is small relative to the parent width, the
encounter region remains nearly uniformly populated and the same
area law applies.

\begin{figure*}[t]
 \centering
 \includegraphics[width=\textwidth]{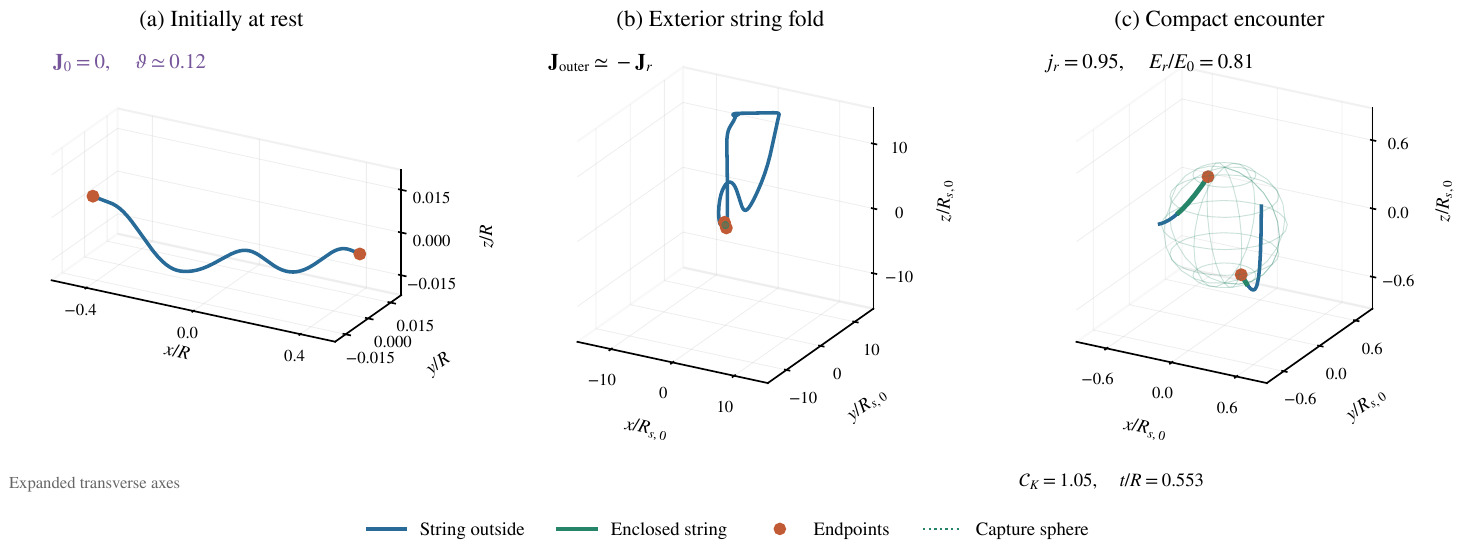}
 \caption{Spin from an initially stationary nonplanar string,
 with \(\vartheta=0.12\) and \(\alpha_{\rm gr}=3\times10^{-4}\).
 (a) Initial shape, with expanded transverse axes.
 (b,c) The exterior string fold and enlarged compact encounter at the same time.
 The core contains 81 percent of the energy and has \(j_r=0.95\);
 the outer string carries the opposite angular momentum.
 Here \(R_{s,0}=2GE_0\), and the sphere marks
 Eq.~\eqref{eq:criterion}. Evolution stops at capture.}
 \label{sup:fig:shapeonly}
\end{figure*}

\textit{Shape alone.}
\label{sup:shapeonly}
Figure~\ref{sup:fig:shapeonly} shows a stationary nonplanar string forming
a core with \(j_r\simeq0.95\), despite zero total initial angular momentum.
The shape-only sample contains over 16,000 trajectories. At
\(\alpha_{\rm gr}=3\times10^{-4}\), the
\(\vartheta=0.12\) and 0.16 samples give 811 captures from 6912 trials
and 176 from 6528. Their mean spins are 0.654 and 0.647, consistent
with \(2/3\) within the uncertainty from the finite number of sampled
shapes. Their cumulative spin distributions differ from
\(F(j_r)=j_r^2\) by at most 0.038 and 0.063, respectively.
The endpoints carry more than 99 percent of the captured
energy, even though part of the string remains outside.

Thus a broad distribution of encounter offsets can recover the linear
law without initial rotation. Small bends instead give common,
low-spin capture. These are core-spin statements: absorption of the
entire isolated zero-\(J\) source without angular momentum escaping
would leave a nonrotating hole, as discussed in the Letter.

Shape can also bring an initially missing pair to capture on a later
encounter, without radiation. At \(\alpha_{\rm gr}=10^{-4}\), captures
on the second and third encounters have mean spin 0.66, again consistent
with \(2/3\). Here the changing shape brings the particles together.
Their spin need not approach extremality as it does during radiative
spin-down of an initially super-Kerr source.

\section{Tidal motion and inflationary bending}
\label{sup:selection}

\textit{Tides.}
A differential acceleration \(\Phi/R\) acting for time \(R\) gives
relative transverse speed \(\sim\Phi\) and displacement \(\sim\Phi R\).
This yields Eq.~\eqref{eq:tidalwidth} and coherent tilt
\(\vartheta\sim\Phi\). For a static straight string, the leading
transverse force cancels between energy and tension when the scalar
metric potentials are equal~\cite{DeVilliersFrolov1998}. The massive
endpoints still experience differential acceleration, which torques
the pair. Their motion then launches transverse waves along the string.

We check this response using
\(\Phi_{\rm ext}=\mathcal T_y(\tau)xy+\mathcal T_z(\tau)xz\),
with the pair initially along \(x\) and tidal-tensor components
\(\mathcal T_{y,z}\sim\Phi/R^2\). We test smooth tidal fields that vary
in strength, reverse sign or change direction. The field is switched off
at \(\tau=0.3R\), after which the string contracts freely.
The forcing is first order in the external field, while
the subsequent evolution uses the nonlinear string constraints.
Twelve tidal-only and six mixed tidal--bend histories, spanning
\(\mu R/m=6,18,60,180\), give 54 captures in 54 sub-Kerr tests and none
in 36 super-Kerr tests. Tidal-only captures enclose the full sampled
source. Added bends leave at most 0.44 percent of its energy outside
and shift the core spin by at most 3.4 percent.

\textit{Inflation.}
For a pre-existing string of constant tension, the weak-bend
spectrum of frozen transverse fluctuations is
\(|y_k|^2\sim H_{\rm inf}^2/(\mu k^3)\), where \(y_k\) and \(k\) are
comoving displacement and wavenumber~\cite{GarrigaVilenkin1993,StephanOtto2004}.
A physical displacement \(\delta Y_\perp\) at physical wavenumber
\(k_\parallel\) has tilt \(\sim k_\parallel\delta Y_\perp\).
Integrating the one-dimensional spectrum gives a pair-scale tilt
\begin{equation}
 \vartheta_{\rm inf}\sim\frac{H_{\rm inf}}{\sqrt\mu}
       \sim\sqrt{\frac{\epsilon_{\rm sr}}{\alpha_{\rm gr}}}\,\Phi.
 \label{sup:slowroll}
\end{equation}
Here \(\epsilon_{\rm sr}=-d\ln H_{\rm inf}/d\ln a_{\rm FRW}\), with
\(a_{\rm FRW}\) the cosmological scale factor. The second relation uses
\(\Phi\sim H_{\rm inf}\sqrt{G/\epsilon_{\rm sr}}\) on the same
scale~\cite{LythRiotto1999}, suppressing numerical and transfer factors. Equation~\eqref{eq:window}
then becomes \(\alpha_{\rm gr}\gg\sqrt{\epsilon_{\rm sr}}\Phi\), or
\(\alpha_{\rm gr}\gg H_{\rm inf}\sqrt G\).

At constant tension, the small-bend condition implies
\(\mu/H_{\rm inf}^2\gg1/(H_{\rm inf}\sqrt G)\gg1\). Spontaneous
de Sitter loop nucleation is then exponentially suppressed, with a rate
proportional to \(\exp[-4\pi\mu/H_{\rm inf}^2]\)~\cite{GarrigaVilenkin1993}.
Efficient nucleation by this process therefore favors larger bends.

For constant tension, the linear string mode satisfies
\begin{equation}
 \partial_\eta(a_{\rm FRW}^2\partial_\eta y_k)
                 +a_{\rm FRW}^2k^2y_k=0,
 \label{sup:frozenmode}
\end{equation}
where \(\eta\) is conformal time~\cite{StephanOtto2004}.
Outside the horizon, the leading shape is constant and the independent
bulk velocity mode decays as \(\partial_\eta y_k\propto a_{\rm FRW}^{-2}\).
Re-entry sets the bend moving without generating total angular momentum
in an initially nonrotating isolated pair and string.

Shorter weak modes reenter earlier. Their physical amplitudes remain
approximately constant while their wavelengths grow, reducing their
tilts~\cite{StephanOtto2004}. For radiation-era re-entry of both modes and
pair, their mean-square contribution at pair entry is
\begin{equation}
 \vartheta_{\rm rms,short}^2\sim\frac{H_{\rm inf}^2}{\mu}
       \int_{n_{\min}}^{n_{\max}}\frac{d\ln n}{n^2},
 \label{sup:uv}
\end{equation}
with \(n\) the wavenumber relative to the pair scale.

\section{Radiation and other losses}
\label{sup:radiation}

\textit{Waveform and spin-down.}
We calculate radiation from the complete conserved source. In the gauge
of Eq.~\eqref{sup:bulk}, its spatial stress is~\cite{Martin:1996cp}
\begin{align}
 T^{ij}_{\rm s}&=\mu\int d\sigma\,
  (\dot X^i\dot X^j-X^{\prime i}X^{\prime j})
             \delta^3(\mathbf x-\mathbf X),\nonumber\\
 T^{ij}_{\rm e}&=\sum_a m\gamma_a v_a^iv_a^j
             \delta^3(\mathbf x-\mathbf x_a).
 \label{sup:stress}
\end{align}
For observation direction
\(\mathbf n\) and retarded time \(u\), define
\begin{equation}
 \mathcal H(u,\mathbf n)=(e_+^{ij}-ie_\times^{ij})
 \int d^3x\,T_{ij}(u+\mathbf n\cdot\mathbf x,\mathbf x).
 \label{sup:waveform}
\end{equation}
The polarization tensors have norm \(e_A^{ij}e_{B\,ij}=2\delta_{AB}\);
the strain at distance \(D\) is \(h_+-ih_\times=(2G/D)\mathcal H\).
The outward energy and angular-momentum fluxes are~\cite{Martin:1996cp,Ruiz2008}
\begin{align}
 P_{\rm GW}&=\frac{G}{4\pi}\int d\Omega\,|\partial_u\mathcal H|^2,\nonumber\\
 L_z&=-\frac{G}{4\pi}\operatorname{Re}\int d\Omega\,
       (\partial_\phi\mathcal H)(\partial_u\mathcal H)^*.
 \label{sup:flux}
\end{align}
Here \(z\) is the angular-momentum axis and the polarizations use the
spherical \((\theta,\phi)\) basis. Integrating over a cycle gives
\(\Delta E\) and the parallel angular-momentum loss \(\Delta J\).
A mode \(e^{-i\omega u+im_\phi\phi}\) carries
\(\Delta J_z/\Delta E=m_\phi/\omega\), so its frequency and angular
structure both matter. The normalization reproduces the straight
oscillator~\cite{Martin:1996cp} and the rigid-rotation relation
\(\Delta J/\Delta E=1/\Omega\)~\cite{Babichev2005}, with rotation rate
\(\Omega\), to about one part in \(10^5\) for the latter.
Field-theory simulations also recover the point-source radiation spectrum
before the monopoles collide~\cite{DvaliConfinedMonopoles2023}.

For a rotating source, define
\begin{equation}
 e=\frac{\Delta E}{E},\qquad
 \mathcal R=\frac{E\Delta J}{2J\Delta E},\qquad
 \frac{j_{\rm next}}{j}=\frac{1-2\mathcal R e}{(1-e)^2}.
 \label{sup:loss}
\end{equation}
The perpendicular torque is small in these examples. Thus weak losses
lower \(j\) when \(\mathcal R>1\). Writing the power as
\(\Gamma G\mu^2\), with dimensionless radiation efficiency \(\Gamma\),
a period of order \(R\) gives \(e\sim\Gamma\alpha_{\rm gr}\).
Between encounters, the string evolves through the Nambu--Goto equations.
At each turning point, we retain its evolved shape and motion, adjusting
its size and velocity amplitude to account for the energy and angular
momentum radiated during the previous cycle. We then recalculate the
waveform. This gives an osculating evolution for weak losses.
The pairs that reach capture do so close to extremality. Shape evolution
changes whether and when capture occurs, and therefore the mass retained
by the hole.

The same energy-loss estimate allows
\(N_{\rm rad}\sim e^{-1}\sim(\Gamma\alpha_{\rm gr})^{-1}\) oscillations
before an order-one energy loss. For shape-dominated pairs that survive,
this can enhance the small
per-passage capture probability by the same factor. The order-unity-bend
estimate \(\sim\alpha_{\rm gr}^2\) in the Letter would then give a
cumulative probability \(\sim\alpha_{\rm gr}/\Gamma\), while capture
remains rare.

\phantomsection
\textit{Background from super-Kerr capture.}
\label{sup:background}
For the background estimate, we take pairs with \(1<j_0<1.5\) to
collapse after radiating. The area law gives roughly as many such pairs
as prompt candidates with \(j_0<1\). We use an illustrative radiated
fraction \(\epsilon\sim0.1\), giving
\(M_{\rm BH}\simeq(1-\epsilon)E_0\), where \(E_0\) is the initial
energy. At power \(\Gamma G\mu^2\),
\begin{equation}
 \Delta t\sim\frac{\epsilon GE_0}{\Gamma\alpha_{\rm gr}^2},
 \qquad
 \Omega_{\rm GW,0}\sim\Omega_{\rm DM,0}f_{\rm PBH}^{\rm delayed}
        \frac{\epsilon}{1-\epsilon}\frac{a_{\rm em}}{a_0}.
 \label{sup:gwenergy}
\end{equation}
The second relation redshifts the emitted energy relative to the surviving
remnant density~\cite{Phinney:2001di}. Here \(\Omega_{\rm DM,0}\) is today's
dark-matter density in critical units, \(f_{\rm PBH}^{\rm delayed}\) its
fraction in delayed remnants, and \(a_{\rm em}/a_0\) the scale-factor
ratio near the end of emission. Energy in an exterior string fold is not
included in \(\epsilon\).

For emission lasting well beyond horizon entry and ending during
radiation domination, \(a_{\rm em}/a_0\propto E_0^{1/2}/\alpha_{\rm gr}\).
Its mass dependence cancels that in Eq.~\eqref{eq:abundance}.

\phantomsection
\textit{Plasma losses.}
\label{sup:plasma}
For charged endpoints in a weakly coupled relativistic bath of temperature
\(T\), let \(\alpha_{\rm EM}=e_{\rm EM}^2/(4\pi)\).
The couplings and particle masses entering these estimates may differ
from their present values as scalar backgrounds, including moduli,
evolve~\cite{Dvali:1995ce,Dvali:2021byy}.
The loss conditions below must hold during the infall leading to collapse,
with the bath properties, couplings and masses evaluated at that epoch.
A stopping force \(F\)
acting over distance \(R\) removes a fraction \(F/\mu\) of the stored
energy. The leading estimates are~\cite{BraatenThoma1991,PeigneSmilga2009,BerezinskyMartinVilenkin1997}
\begin{equation}
 \begin{gathered}
 \frac{\Delta E_{\rm coll}}{E}\sim\frac{\alpha_{\rm EM}^2T^2}{\mu},
 \qquad \frac{\Delta E_{\rm acc}}{E}\sim\frac{\alpha_{\rm EM}\mu}{m^2},\\
 \frac{\Delta E_{\rm brem}}{E}\sim
              \frac{\alpha_{\rm EM}^3T^3R}{m^2}.
 \end{gathered}
 \label{sup:plasma-loss}
\end{equation}
The subscripts ``coll'', ``acc'' and ``brem'' denote collisional stopping
in the plasma, radiation from string-driven acceleration and
scattering-induced bremsstrahlung, respectively.
Charge, species and logarithmic factors are suppressed.
The bremsstrahlung estimate assumes \(m^4\gg\alpha_{\rm EM}^2T^3E\)
and \(m^2\gg\mu\),
so the endpoint mass sets the photon-production timescale, which is short
compared with the acceleration timescale. The first two losses are small for
\(T\lesssim\sqrt\mu\ll m\ll\mu R\), while bremsstrahlung also constrains
\(R\). Plasma kicks must leave the selected angular momentum nearly
unchanged. Repeated encounters additionally require these losses and any
monopole plasma damping or gauge radiation to remain below gravitational
emission~\cite{BaisLangacker1982,DunskyEtAl2022,BerezinskyMartinVilenkin1997}.

\end{document}